\documentclass[12pt,reqno]{article}
\usepackage{amsfonts}
\usepackage{amsmath}
\usepackage{amsbsy}
\usepackage{graphicx}
\usepackage{amssymb,latexsym}
 \numberwithin{equation}{section}

\begin{document}
 \allowdisplaybreaks[1]
\title{Effective Matter Cosmologies of Massive Gravity: Physical Fluids}
\author{Nejat Tevfik Y$\i$lmaz\\
Department of Electrical and Electronics Engineering,\\
Ya\c{s}ar University,\\
Sel\c{c}uk Ya\c{s}ar Kamp\"{u}s\"{u}\\
\"{U}niversite Caddesi, No:35-37,\\
A\u{g}a\c{c}l\i Yol, 35100,\\
Bornova, \.{I}zmir, Turkey.\\
\texttt{nejat.yilmaz@yasar.edu.tr}} \maketitle
\begin{abstract}
We derive new cosmological solutions of the ghost-free massive
gravity with a general background metric in which the contribution
of the mass sector to the metric one is modeled by an effective
cosmological constant and an ideal fluid which obeys the first law
of thermodynamics; thus it satisfies the ordinary energy-momentum
conservation or continuity equation.
\end{abstract}

\section{Introduction}

In the following work we construct a new class of
Friedmann-Lemaitre-Robertson-Walker (FLRW) cosmological solutions
of the ghost-free \cite{BD1,BD2} massive gravity theory
\cite{dgrt1,dgrt2,hr1,hr2,hr3} which is a nonlinearized version of
the Fierz-Pauli \cite{fp} massive gravity. The cosmological
solutions of the ghost-free massive gravity have been extensively
studied in recent years. It has already been shown that for a flat
background metric there exist open FLRW cosmological solutions
\cite{gum} but no flat or closed ones \cite{mc}. Hence the program
has evolved to search the cosmological solutions for the de Sitter
\cite{ln} and the FLRW type \cite{higuchi} background metrics. The
underlying motivation to construct the cosmological solutions of
ghost-free massive gravity in these as well as other works
\cite{koyama1,koyama2,koyama3,koyama4,volkov1,volkov2,ih1} is to
obtain phenomenologically acceptable self-accelerating solutions
which will admit self-acceleration without a need for a
matter-generated cosmological constant. However, the solutions for
the above-mentioned natural background metrics have been shown to
posses stability problems \cite{higuchi,sta1,derham} which turn
them into physically unacceptable candidates. One way to overcome
the stability problem in the cosmological solutions is to give up
the homogeneity and, or the isotropy of the physical metric. In
that case one may recover the FLRW cosmologies at the regime of
the Compton wavelength of the mass if it is of the order of the
Hubble constant \cite{inhom}. Another route is to consider
nonstandard background metrics for which the stability and
symmetry considerations of the cosmological solutions are still
the primary concern for the physical qualification of the
solution.

In the general literature of deriving cosmological solutions of
massive gravity there exist two technical approaches. One of them
is to fix the background metric as a flat one and to solve the
field equations upon introducing an ansatz for the St\"{u}ckelberg
fields. This has been the mainstream treatment which (as we have
discussed above) results in physically problematic solutions. The
other approach is to introduce a background metric ansatz in terms
of the St\"{u}ckelberg scalars whose forms are also specially
chosen. In this case by substituting the ansatz into the scalar
field equations (which are derived by varying the gravitational
potential coming from the mass terms) or by using the equivalent
conservation equation of the mass contribution to the metric
equation \cite{volkov1,volkov2} one obtains the particular scalar
solutions and thus the accompanying background metric for these
solutions. Then one is able to construct the associated
cosmological equations whose solutions are under consideration. In
defining the mass contribution to the metric sector, usually the
effective energy-momentum tensor is introduced. In particular,
when one considers the cosmological solutions this leads to an
effective ideal fluid contribution to the cosmological dynamics
apart from the matter presence. However, a great majority of the
literature is restricted to only a cosmological constant
contribution via the effective fluid for the purpose of achieving
self-accelerating solutions.

In the present paper instead of assigning a particular ansatz for
the background metric in terms of the St\"{u}ckelberg scalars (for
example, like the case in Ref. \cite{ih1}) our technical objective
will be to derive its functional form constructed from the scalar
and the physical metric solutions when it is diagonal with the
entries that are functions of a single coordinate, so that the
field equations are satisfied with such a choice of background
metric. By focusing on the physical FLRW metrics we will be able
to derive an extensive class of exact FLRW cosmological solutions
of massive gravity for a functionally parametrized set of general
background metrics for which the St\"{u}ckelberg dependence is
implicit. Therefore, in the following analysis our main concern
will be to design the background metrics necessary to construct
exact FLRW solutions of the theory. This approach is quite
different than the ones used in other treatments (referred to
above), which either used a flat background metric or introduced a
particular St\"{u}ckelberg dependence and solved the scalar and
cosmological metric solutions thereafter. On the contrary in the
following original solution construction technique which starts
from an arbitrary background metric by considering the field
equations which couple it to the scalar and the gravity sectors of
the theory we will devise a method to find its functional form in
terms of the St\"{u}ckelberg-sector solutions, and the FLRW
physical metric that is necessary to satisfy all the nonlinear
field equations of the theory. Our methodology in this direction
will be in parallel with that used in Ref. \cite{cosmommgr} (where
a class of solutions of the minimal ghost-free massive gravity
model \cite{hr1,hr2,hr3} were constructed), and it will follow the
same mathematical track as in Ref. \cite{massgrav}. Within this
general approach one finds the means to decouple the scalar and
gravity field equations. Essentially the decoupling follows an
identification or a definition of the contribution coming from the
St\"{u}ckelberg scalar fields to the metric equation as an
effective cosmological constant and an energy-momentum tensor
which is well known in the literature. Since the covariant
constancy of this contribution is equivalent to the scalar field
equations from the above-mentioned identification, one can
transform the scalar equations to a constraint condition on the
effective energy-momentum tensor. Therefore, one may group all the
dynamical equations of the theory in the metric sector in which an
effective matter contribution is included. Such a compactified and
scalar-free way of writing the metric equation results in the
well-known Einstein form in the presence of some effective matter
which obeys a conservation equation that is equivalent to the
dynamical equations of the scalars. Consequently in this respect
one decouples the scalar and metric sectors by collecting all the
dynamics in the metric sector and singling out an algebraic
equation, which must be satisfied by the background and physical
metrics as well as the scalars and the effective matter (which
originally stems from the definition of the effective cosmological
constant and the energy-momentum tensor contribution of the
St\"{u}ckelberg fields). Owing to its role in this decoupling
mechanism of the field equations, we call this algebraic equation
the solution ansatz in the following treatment. With this
underlying general framework the main purpose of this paper will
be to focus on the homogeneous and isotropic cosmological
solutions of the theory. Hence, in this specialization apart from
using the FLRW physical metric we will also consider an
ideal-fluid form for the effective energy-momentum tensor of the
solution ansatz. In this case, performing the above-mentioned
decoupling will lead us to the standard cosmological dynamics
whose well-known solutions depend on the equation of state of the
ideal effective fluid (which in our derivation is completely
arbitrary) to generate and parametrize the solutions. We we will
also derive the diagonal solutions of the algebraic ansatz
equation for the background metric in terms of the energy density
and the pressure of the effective fluid, the scale factor of the
FLRW metric, and the St\"{u}ckelberg scalar fields. In that regard
we will be able to show that when one specifies the
St\"{u}ckelberg scalar fields arbitrarily and solves the effective
energy density, the effective pressure, and the scale factor from
the completely decoupled cosmological sector, one can construct
the necessary background metrics which accompany these exact FLRW
solutions of massive gravity when satisfying the field equations.

As we have remarked above, a similar analysis in Ref.
\cite{massgrav} was used to derive another class of exact FLRW
cosmological solutions of massive gravity. However, there a
different and a more involved solution ansatz was at work which
led to a nonstandard and a nonphysical conservation law for the
effective ideal fluid arising via the application of the covariant
constancy constraint on the ansatz equation (which is equivalent
to the scalar field equations). Thus the resulting nonstandard
continuity equation suggests that the effective ideal fluid which
is a pseudo-ontological ingredient of the theory that takes the
role of summarizing the collective effects of the mass degrees of
freedom in the physical metric sector behaves (unlike the ordinary
matter) nonphysically. The notion of nonphysicality here is due to
the fact that such a conservation law cannot be consistent with
the first law of thermodynamics, which would be the case for a
fluid which exhibits a usual energy-momentum conservation law. In
the present paper on the other hand by staying in a similar
solution scheme our main concern will be to construct completely
new exact FLRW solutions of the theory which are generated by a
somewhat simpler ansatz, which leads to solutions via
physical-like effective fluids that have a standard
energy-momentum conservation law that is compatible with the first
law of thermodynamics.
\section{The Cosmology of Physical Effective Fluids}
 In the following analysis our framework will be the general massive gravity action \cite{hr1}
\begin{equation}\label{e1}
 S=-M_p^2\int\bigg[ R\ast 1-2m^2\sum\limits_{n=\texttt{0}}^{\texttt{3}}\beta_{n} e_{n}(\sqrt{\Sigma})\ast 1+\Lambda\ast
 1\bigg]-S_{matt},
\end{equation}
 for generic coefficients $\beta_n$. The mass terms arise from the elementary
 symmetric polynomials
\begin{subequations}\label{e2}
\begin{align}
e_{0}\equiv e_{0}(\sqrt{\Sigma})&=1,\notag\\
e_{1}\equiv e_{1}(\sqrt{\Sigma})&=tr\sqrt{\Sigma},\notag\\
e_{2}\equiv e_{2}(\sqrt{\Sigma})&=\frac{1}{2}\big((tr\sqrt{\Sigma})^2-tr(\sqrt{\Sigma})^2\big),\notag\\
e_{3}\equiv
e_{3}(\sqrt{\Sigma})&=\frac{1}{6}\big((tr\sqrt{\Sigma})^3-3\,tr\sqrt{\Sigma}\:tr(\sqrt{\Sigma})^2+2\,tr(\sqrt{\Sigma})^3\big)
\tag{\ref{e2}}
\end{align}
\end{subequations}
of the square root matrix $\sqrt{\Sigma}$ of
\begin{equation}\label{e3}
(\Sigma)^{\mu}_{\:\:\:\nu}=g^{\mu\rho}\partial_{\rho}\phi^a\partial_{\nu}\phi^{b}\bar{f}_{ab},
\end{equation}
which couples the inverse physical metric $g^{\mu\nu}$ to the
fiducial one $\bar{f}_{ab}(\phi^{c})$ via the four St\"{u}ckelberg
scalar fields $\{\phi^{a}(x^\mu)\}$. Now if one varies the action
\eqref{e1} with respect to the physical metric one obtains the
metric equation \cite{hr1}
\begin{subequations}\label{e4}
\begin{align}
R_{\mu\nu}&-\frac{1}{2}g_{\mu\nu}R-\frac{1}{2}\Lambda
g_{\mu\nu}+\frac{1}{2}m^2\bigg[\sum\limits_{n=\texttt{0}}^{\texttt{3}}(-1)^n\beta_{n}\bigg(g_{\mu\lambda}Y_{n\,\nu}^{\lambda}(\sqrt{\Sigma})\notag\\
&+g_{\nu\lambda}Y_{n\,\mu}^{\lambda}(\sqrt{\Sigma})\bigg)\bigg]=G_NT^{matt}_{\mu\nu}\tag{\ref{e4}},
\end{align}
\end{subequations}
where
\begin{subequations}\label{e6}
\begin{align}
Y_{0}(\sqrt{\Sigma})&=\mathbf{1}_4,\notag\\
Y_{1}(\sqrt{\Sigma})&=\sqrt{\Sigma}-tr\sqrt{\Sigma}\,\mathbf{1}_4,\notag\\
Y_{2}(\sqrt{\Sigma})&=(\sqrt{\Sigma})^2-tr\sqrt{\Sigma}\sqrt{\Sigma}+\frac{1}{2}\big[(tr\sqrt{\Sigma})^2-tr(\sqrt{\Sigma})^2\big]\mathbf{1}_4,\notag\\
Y_{3}(\sqrt{\Sigma})&=(\sqrt{\Sigma})^3-tr\sqrt{\Sigma}\,(\sqrt{\Sigma})^2+\frac{1}{2}\big[(tr\sqrt{\Sigma})^2-tr(\sqrt{\Sigma})^2\big]\sqrt{\Sigma}\notag\\
&-\frac{1}{6}\big[(tr\sqrt{\Sigma})^3-3\,tr\sqrt{\Sigma}\:tr(\sqrt{\Sigma})^2+2\,tr(\sqrt{\Sigma})^3\big]\mathbf{1}_4,
\tag{\ref{e6}}
\end{align}
\end{subequations}
where all terms are $4\, \times 4\,$ matrices. On the other hand,
the field equations of Eq.\eqref{e1} for the St\"{u}ckelberg
scalar fields $\{\phi^{a}(x^\mu)\}$ can equivalently be written as
\begin{equation}\label{e6.1}
\nabla^{\mu}\bigg[\sum\limits_{n=\texttt{0}}^{\texttt{3}}(-1)^n\beta_{n}\bigg(g_{\mu\lambda}Y_{n\,\nu}^{\lambda}(\sqrt{\Sigma})
+g_{\nu\lambda}Y_{n\,\mu}^{\lambda}(\sqrt{\Sigma})\bigg)\bigg]=0.
\end{equation}
Now let us consider the homogeneous and isotropic solutions of
Eq.\eqref{e4}. For this reason we take the physical metric to be
the FLRW one,
\begin{equation}\label{e7}
g=-dt^2+\frac{a^2(t)}{1-kr^2}dr^2+a^2(t)r^2 d\theta^2 +a^2(t)r^2
sin^2 \theta d\varphi^2,
\end{equation}
and the physical matter as a perfect fluid whose energy-momentum
tensor reads
\begin{equation}\label{e8}
T^{matt}_{\mu\nu}=(\rho+p)U_{\mu}U_{\nu}+pg_{\mu\nu},
\end{equation}
with $p=p(t)$ and $\rho=\rho(t)$ being the pressure and energy
density of the fluid, respectively. We take the fluid
four-velocity vector as $U_{\mu}=(1,0,0,0)$ which is defined in
the rest frame of the fluid. Aiming to derive the cosmological
dynamics that admit solutions of Eq.\eqref{e4} in the form
\eqref{e7}, we propose the solution ansatz
\begin{equation}\label{e9}
\frac{1}{2}m^2\bigg[\sum\limits_{n=\texttt{0}}^{\texttt{3}}(-1)^n\beta_{n}\bigg(g_{\mu\lambda}Y_{n\,\nu}^{\lambda}(\sqrt{\Sigma})
+g_{\nu\lambda}Y_{n\,\mu}^{\lambda}(\sqrt{\Sigma})\bigg)\bigg]=C_1m^2g_{\mu\nu}+C_2m^2\tilde{T}^{eff}_{\mu\nu},
\end{equation}
where $C_1$ and $C_2$ are arbitrary constants and we assume that
$\tilde{T}^{eff}$ is also of the ideal fluid form,\footnote{For
future convenience here we define the matrix
$[\tilde{T}^{eff}]^\mu_{\:\:\:\nu}:=\tilde{T}^{eff}_{\mu\nu}$.}
\begin{equation}\label{e10}
\tilde{T}^{eff}_{\mu\nu}=(\tilde{\rho}(t)+\tilde{p}(t))U_{\mu}U_{\nu}+\tilde{p}(t)g_{\mu\nu}.
\end{equation}
At this point we should emphasize once more that both in Ref.
\cite{massgrav} and here, following the introduction of an
effective cosmological constant and a fluid in the solution ansatz
the perspective is to derive the necessary background metric that
will lead us to the solutions of the theory by solving the
St\"{u}ckelberg-sector fields and by constructing the cosmological
dynamics ready to be solved for the scale factor, effective energy
density, and pressure of the hypothetical fluid. In Ref.
\cite{massgrav}, however, the solutions were based on the
assumption of an effective fluid that exhibits a nonphysical
conservation law thus does not obey the first law of
thermodynamics. In the present treatment on the other hand by
being in a similar solution scheme as that in Ref. \cite{massgrav}
we will be aiming to obtain completely new solutions of the theory
which are generated by a physical-like effective fluid used in the
ansatz which obeys the first law of thermodynamics. Therefore the
solution ansatz differs from the one we employed in Ref.
\cite{massgrav}, whose unique form was dictated by the nonphysical
nature of the corresponding effective fluid we considered. The
nonphysical nature was due to the fact that the effective fluid
Lagrangian in Ref. \cite{massgrav} was chosen in its most general
form so that one does not have to use the first law of
thermodynamics to derive the energy-momentum tensor \eqref{e10}
upon varying the Lagrangian with respect to the inverse metric. In
this work we will consider the physically oriented effective
fluids which obey the first law of thermodynamics,
\begin{equation}\label{e11}
d\tilde{\rho}=\tilde{\mu}d\tilde{n}+\tilde{n}\tilde{T}d\tilde{s},
\end{equation}
where $\tilde{\mu}=(\tilde{\rho}+\tilde{p})/\tilde{n}, \tilde{n},
\tilde{T}, \tilde{s}$ are the chemical potential, the particle
number density, the temperature, and the entropy per particle of
the effective fluid respectively. If one defines the particle
number flux density as
\begin{equation}\label{e12}
J^\mu=\tilde{n}\sqrt{-g}\,U^\mu,
\end{equation}
and introduces the Lagrangian coordinate scalars $\alpha^A$, as
well as the Clebsch scalar potentials $\theta_1,\theta_2,\beta_A$
where $A=1,2,3,$ one can write the ideal fluid action \cite{brown}
as
\begin{equation}\label{e13}
S_{IF}=\int
dx^4\sqrt{-g}\big[-\tilde{\rho}+\frac{1}{\sqrt{-g}}\;J^\mu\big(\partial_\mu\theta_1+\tilde{s}\partial_\mu\theta_2+\beta_A\partial_\mu\alpha^A\big)\big].
\end{equation}
Bearing in mind the equation of state
$\tilde{\rho}=\tilde{\rho}(\tilde{n},\tilde{s})$ and the fact that
via Eq.\eqref{e12} we have
\begin{equation}\label{e14}
\tilde{n}=\frac{\sqrt{-J^\mu J_\mu}}{\sqrt{-g}},
\end{equation}
the linearly independent fields composing the action functional
\eqref{e13} become
$\{g^{\mu\nu},J^\mu,\tilde{s},\alpha^A,\theta_1,\theta_2,\beta_A\}$.
By varying this action with respect to the inverse metric, using
the field equations of $J^{\mu}$ [obtained by varying
Eq.\eqref{e13} with respect to $J^{\mu}$], and demanding that the
effective fluid satisfies the first law \eqref{e11}, via
\begin{equation}\label{e15}
\tilde{T}^{eff}_{\mu\nu}=-2\frac{1}{\sqrt{-g}}\frac{\delta(\sqrt{-g}\mathcal{L}_{IF})}{\delta(g^{\mu\nu})},
\end{equation}
one can derive the effective ideal fluid energy-momentum tensor
\eqref{e10}. Hence, we observe that in order for the action
\eqref{e13} to yield Eq.\eqref{e10} the fluid must be constrained
to be a physical one that satisfies the first law of
thermodynamics. One can furthermore show that if the field
equations of the particle number flux densities $J^\mu$ are used
in Eq.\eqref{e13} then the Lagrangian of the physically behaving
effective fluid becomes
\begin{equation}\label{e16}
\mathcal{L}_{IF}=\tilde{p}.
\end{equation}
Putting this result aside, our next task is to identify the
on-shell Lagrangian of the mass terms in Eq.\eqref{e1}, that is
the Lagrangian-level ansatz which generates Eq.\eqref{e9}. From
Eqs.\eqref{e1} and \eqref{e4} for the solutions which satisfy the
ansatz \eqref{e9}, we can deduce that the variation with respect
to the inverse physical metric must yield (on-shell)
\begin{equation}\label{e17}
 \delta\big(2m^2\sqrt{-g}\sum\limits_{n=\texttt{0}}^{\texttt{3}}\beta_{n}
 e_{n}\big)=-\sqrt{-g}\big[C_1m^2g_{\mu\nu}+C_2m^2\tilde{T}^{eff}_{\mu\nu}\big]\delta g^{\mu\nu}.
\end{equation}
Therefore from our above discussion we conclude that for the
solutions satisfying Eq.\eqref{e9} we must have the on-shell
identity
\begin{equation}\label{e17.1}
 \sum\limits_{n=\texttt{0}}^{\texttt{3}}\beta_{n}
 e_{n}=C_1+C_2\tilde{p}.
\end{equation}
We should remark once more that for this on-shell identity to hold
we must demand that the effective ideal fluid whose
energy-momentum tensor is used in the solution ansatz \eqref{e9}
must satisfy the first law of thermodynamics \eqref{e11}. For this
reason it deserves to be called physical. Thus in order to be able
to use the solution \eqref{e9} and the corresponding Lagrangian
ansatz \eqref{e17} together we must introduce Eq.\eqref{e11} as a
mathematical constraint on the effective fluid. On the other hand
to satisfy the St\"{u}ckelberg scalar field equations \eqref{e6.1}
we observe that our solution ansatz \eqref{e9} demands that the
equation
\begin{equation}\label{e18}
\nabla^\mu\big[C_1g_{\mu\nu}+C_2\tilde{T}^{eff}_{\mu\nu}\big]=0,
\end{equation}
must be satisfied by the solution-generating-and-parametrizing
fields composing $\tilde{T}^{eff}$. By using the metric
compatibility of the spin connection this constraint leads us to
\begin{equation}\label{e19}
\nabla^\mu\tilde{T}^{eff}_{\mu\nu}=0,
\end{equation}
which is in the form of the usual continuity or energy-momentum
conservation equation for a physical fluid which we would expect
to see following our assumption of a physical-like behavior for
our solution-parametrizing effective fluid via demanding that it
should satisfy Eq.\eqref{e11}. Hence unlike the case studied in
Ref. \cite{massgrav}, we get no modification to the fluid equation
of the effective ideal fluid in this solution scheme. Our next
task is to find the St\"{u}ckelberg fields and the background
metric in terms of the physical metric, the coefficients
$C_{1},C_{2}$, and the pressure and the energy density of the
effective fluid that satisfy the solution ansatz \eqref{e9}. Now,
by using the matrix identity
$g(\sqrt{\Sigma})^n=(g(\sqrt{\Sigma})^n)^T$, for any integer $n$
\cite{bac}, as well as the definitions of the elementary symmetric
polynomials in Eq.\eqref{e2}, and the constraint \eqref{e17.1}
(which attributes a physical nature to the effective fluid
introduced to parametrize the solution ansatz), from Eq.\eqref{e9}
(after some algebra) we obtain the algebraic matrix
equation\footnote{The reader may refer to Ref. \cite{massgrav} for
the details of a similar computation.}
\begin{subequations}\label{e20}
\begin{align}
&-\beta_3(\sqrt{\Sigma})^3+\big(\beta_2+\beta_3e_1\big)(\sqrt{\Sigma})^2+\big(-\beta_1-\beta_2e_1-\beta_3e_2\big)
(\sqrt{\Sigma})\notag\\
\notag\\
&+C_2\tilde{p}\mathbf{1}_4-C_2g^{-1}\tilde{T}^{eff}=0.\tag{\ref{e20}}
\end{align}
\end{subequations}
If we take the trace of this equation by using Eqs.\eqref{e2} and
\eqref{e17.1} we can write $e_2$ in terms of $e_1$. It reads
\begin{equation}\label{e21}
 e_{2}=\frac{1}{\beta_2}\big[3C_1-3\beta_0-2\beta_1e_1+C_2\tilde{T}^\mu_{\:\:\:\mu}-C_2\tilde{p}\big],
\end{equation}
where we define the contraction
$\tilde{T}^\mu_{\:\:\:\mu}:=g^{\mu\nu}\tilde{T}^{eff}_{\mu\nu}$.
This relation reduces the number of elementary symmetric
polynomial functions to one; namely, in our solution scheme $e_1$
remains to be specified by the particular solution chosen. A
subset of the general solutions of Eq.\eqref{e20} with a different
constant coefficient term were derived in Ref. \cite{massgrav}
under the assumption of the diagonality of $\sqrt{\Sigma}$. If we
adopt those solutions from Ref. \cite{massgrav} while bearing in
mind the variant constant term in Eq.\eqref{e20}, then the
background metric solutions can be given as
\begin{equation}\label{e21.44}
\bar{f}=diag\bigg(\frac{\mathcal{V}^{\prime}_{00}}{\big(F_0(x^0)\big)^2},\frac{\mathcal{V}^{\prime}_{11}}{\big(F_1(x^1)\big)^2}
,\frac{\mathcal{V}^{\prime}_{22}}{\big(F_2(x^2)\big)^2},\frac{\mathcal{V}^{\prime}_{33}}{\big(F_3(x^3)\big)^2}\bigg),
\end{equation}
where $\{F_{\mu}\}$ are arbitrary functions of only the single
coordinate $x^{\mu}$, and the definitions of the diagonal matrices
$\mathcal{V}^{\prime}$ are given in the Appendix. On the other
hand, the St\"{u}ckelberg scalar fields that solve Eq.\eqref{e20}
become \cite{cosmommgr,massgrav}
\begin{equation}\label{e21.8}
\phi^{c}(x^{c}) =\pm \int F_{c}(x^{c})dx^{c}.
\end{equation}
Hence the reader may immediately realize that they are completely
arbitrary. Having found the scalar-sector solutions arising from
the ansatz \eqref{e9}, let us substitute Eq.\eqref{e9} into the
metric equation \eqref{e4} to find the corresponding cosmological
dynamics upon the choice of the FLRW metric \eqref{e7}. We get the
equation
\begin{equation}\label{e22}
R_{\mu\nu}-\frac{1}{2}Rg_{\mu\nu}-\tilde{\Lambda}
g_{\mu\nu}=G_NT^{matt}_{\mu\nu}-C_2m^2\tilde{T}^{eff}_{\mu\nu},
\end{equation}
where we have introduced the effective cosmological constant
$\tilde{\Lambda}=\frac{1}{2}\Lambda-C_1m^2$ which has a
contribution coming from the ansatz \eqref{e9}. By referring to
the ideal-fluid nature of the real and effective matter sources
namely, by using the energy-momentum tensors \eqref{e8} and
\eqref{e10}, the computation of Eq.\eqref{e22} for the metric
\eqref{e7} gives the $t$-component equation
\begin{equation}\label{e23}
\big(\frac{\dot{a}}{a}\big)^2+\frac{k}{a^2}=\frac{G_N}{3}\rho-\frac{C_2m^2}{3}\tilde{\rho}-\frac{\tilde{\Lambda}}{3},
\end{equation}
and the three identical spatial-component equations
\begin{equation}\label{e24}
\frac{2\ddot{a}}{a}=-\big(\frac{\dot{a}}{a}\big)^2-\frac{k}{a^2}-G_Np+C_2m^2\tilde{p}
-\tilde{\Lambda}.
\end{equation}
Now by using Eq.\eqref{e23} in \eqref{e24} we get the cosmic
acceleration equation
\begin{equation}\label{e25}
\frac{\ddot{a}}{a}=-\frac{G_N}{6}\big(3p+\rho\big)+\frac{C_2m^2}{6}\big(3\tilde{p}+\tilde{\rho}\big)-\frac{\tilde{\Lambda}}{3}.
\end{equation}
We realize that the Friedmann [Eq.\eqref{e23}] and acceleration
[Eq.\eqref{e25}] equations are in the canonical form of the
ordinary cosmological ones, and additional contributions to the
cosmological constant, fluid pressure, and energy density come
from the graviton mass sector terms in Eq.\eqref{e1}. The fluid
equation
\begin{equation}\label{e26}
\dot{\rho}=-\frac{3\dot{a}}{a}\big(p+\rho\big),
\end{equation}
of the real-matter ideal fluid \eqref{e8} follows from its
energy-momentum conservation law $\nabla^\mu T^{matt}_{\mu\nu}=0$
which is computed for the FLRW metric choice \eqref{e7}. On the
other hand a similar continuity equation
\begin{equation}\label{e27}
\dot{\tilde{\rho}}=-\frac{3\dot{a}}{a}\big(\tilde{p}+\tilde{\rho}\big),
\end{equation}
which is also in the canonical form must be satisfied by the
effective ideal fluid (which has physical-like properties) via the
presence of the constraint equation \eqref{e19} which has replaced
the St\"{u}ckelberg scalar field equations upon the introduction
of the ansatz \eqref{e9}. Likewise in the standard cosmology case,
Eqs.\eqref{e26} and \eqref{e27} must be solved together with the
Friedmann equation to derive the cosmological solutions. We should
remark here that as a result of our analysis we get a set of
cosmological equations for massive gravity which do not get
dynamical modification terms, i.e., they are in the same form as
the equations of standard cosmology; hence, they exhibit the same
solution moduli. The only extra contribution that appears due to
the presence of the mass sector of the theory is the mathematical
existence of an effective ideal fluid (which lacks any other
physical role apart from the gravitational one, of course), in
addition to the physical content of the Universe. Moreover, the
equation of state of this effective ideal fluid is completely
arbitrary.
\section{Appendix}
 Here we will present the formal details of the
solutions of Eq.\eqref{e20} in parallel with the achievements of
Ref. \cite{massgrav}. By assuming a diagonal form for $g, \Sigma,
\bar{f},$ and $\tilde{T}^{eff}$ we can give the three ($i=1,2,3$)
distinct solutions of the cubic equation \eqref{e20} as
\begin{equation}\label{e28}
\sqrt{\Sigma}=\mathcal{V}_i,
\end{equation}
where
\begin{equation}\label{e29}
\mathcal{V}_i=-\frac{1}{3}\mathbf{a}^{-1}\big[\mathbf{b}\mathbf{1}_4+u_i\mathcal{U}+u_i^{-1}\mathcal{U}^{-1}\big(\mathbf{b}^2-3\mathbf{ac}\big)\big],
\end{equation}
and $\mathbf{1}_4$ is the unit $4\,\times\, 4$ matrix. The
coefficients that differ for each solution are
\begin{equation}\label{e30}
u_1=1,\quad u_2=\frac{1}{2}(-1+i\sqrt{3}),\quad
u_3=\frac{1}{2}(-1-i\sqrt{3}).
\end{equation}
Also the matrix $\mathcal{U}$ is
\begin{equation}\label{e31}
\mathcal{U}=\bigg[\frac{2\mathbf{b}^3-9\mathbf{abc}+27\mathbf{a}^2\mathbf{d}+\sqrt{(2\mathbf{b}^3-9\mathbf{abc}+27\mathbf{a}^2\mathbf{d})^2
-4(\mathbf{b}^2-3\mathbf{ac})^3}}{2}\bigg]^{1/3}.
\end{equation}
Here the constant matrices $\mathbf{a,b,c}$ are the same as those
in Ref. \cite{massgrav}. They are
\begin{subequations}\label{e32}
\begin{align}
\mathbf{a}&=-\beta_3\mathbf{1}_4,\notag\\
\mathbf{b}&=\big(\beta_2+\beta_3e_1\big)\mathbf{1}_4,\notag\\
\mathbf{c}&=\big(-\beta_1-\beta_2e_1-\beta_3e_2\big)\mathbf{1}_4.
\tag{\ref{e32}}
\end{align}
\end{subequations}
However the matrix function $\mathbf{d}$ differs from the
functional form of the one in Ref. \cite{massgrav} it reads
\begin{equation}\label{e33}
\mathbf{d}=C_2\tilde{p}\mathbf{1}_4-C_2g^{-1}\tilde{T}^{eff}.
\end{equation}
In the rest frame of the effective fluid [via Eq.\eqref{e7} and
the effective ideal fluid energy-momentum tensor \eqref{e10}], we
can explicitly calculate $\mathbf{d}$, which becomes
\begin{subequations}\label{e34}
\begin{align}
\mathbf{d}&= C_2\left(\begin{matrix}
\tilde{p}&\texttt{0}&\texttt{0}&\texttt{0}
\\
\texttt{0}&\tilde{p}&\texttt{0}&\texttt{0}\\\texttt{0}&\texttt{0}&\tilde{p}
&\texttt{0}\\\texttt{0}&\texttt{0}
&\texttt{0}&\tilde{p}\end{matrix}\right)-C_2\left(\begin{matrix}
-\texttt{1}&\texttt{0}&\texttt{0}&\texttt{0}
\\
\texttt{0}&g_{\texttt{1}\texttt{1}}^{-1}&\texttt{0}&\texttt{0}\\\texttt{0}&\texttt{0}&g_{\texttt{2}\texttt{2}}^{-1}
&\texttt{0}\\\texttt{0}&\texttt{0}
&\texttt{0}&g_{\texttt{3}\texttt{3}}^{-1}\end{matrix}\right)
\left(\begin{matrix} \tilde{\rho}&\texttt{0}&\texttt{0}&\texttt{0}
\\
\texttt{0}&\tilde{p}g_{\texttt{1}\texttt{1}}&\texttt{0}&\texttt{0}\\\texttt{0}&\texttt{0}&\tilde{p}g_{\texttt{2}\texttt{2}}
&\texttt{0}\\\texttt{0}&\texttt{0}
&\texttt{0}&\tilde{p}g_{\texttt{3}\texttt{3}}\end{matrix}\right)\notag\\\notag\\
&=\left(\begin{matrix}
C_2(\tilde{\rho}+\tilde{p})&\texttt{0}&\texttt{0}&\texttt{0}
\\
\texttt{0}&\texttt{0}&\texttt{0}&\texttt{0}\\\texttt{0}&\texttt{0}&\texttt{0}
&\texttt{0}\\\texttt{0}&\texttt{0}
&\texttt{0}&\texttt{0}\end{matrix}\right)\tag{\ref{e34}}.
\end{align}
\end{subequations}
Similarly, explicit matrix multiplication yields
$\tilde{T}^\mu_{\:\:\:\mu}=3\tilde{p}-\tilde{\rho}$, which appears
in the relation \eqref{e21} that can be used to eliminate $e_2$ in
terms of $e_1$ in the solutions \eqref{e29}. We should state at
this point that within the solutions we have
 constructed, $e_1$ must be specified by the particular choice of solution of Eq.\eqref{e20}. In general, the
solutions given in Eq.\eqref{e29} can be complex; however, we have
to consider the real ones. Now, if we concentrate on the real
solutions in Eq.\eqref{e29} by referring to the definition
\eqref{e3}, Eq.\eqref{e28} can be written as
\begin{equation}\label{e35}
\partial_{\mu}\phi^a\partial_{\nu}\phi^{b}\bar{f}_{ab}=\mathcal{V}^{\prime}_{\mu\nu},
\end{equation}
where we have introduced
\begin{equation}\label{e36}
\mathcal{V}^{\prime}=g\mathcal{V}^2
\end{equation}
and defined the tensor components
\begin{equation}\label{e37}
\mathcal{V}^{\prime}_{\mu\nu}:=
[\mathcal{V}^{\prime}]^{\mu}_{\:\:\:\nu}.
\end{equation}
One can show that \cite{cosmommgr,massgrav,constmmgr} the
background metric choice \eqref{e21.44} and the completely
arbitrary St\"{u}ckelberg scalar field solutions \eqref{e21.8}
presented in Sec. II satisfy the set of first-order partial
differential equations \eqref{e35}. Consequently, Eq.\eqref{e21.8}
forms a solution class of the St\"{u}ckelberg sector of the
massive gravity theory when the background metric is chosen as
Eq.\eqref{e21.44}, whose explicit form depends on the
specification of the effective energy density and the pressure via
the cosmological dynamics. Therefore, in order to generate the
cosmological solutions of the theory under inspection it is
sufficient to solve Eqs.\eqref{e23}, \eqref{e24}, \eqref{e26}, and
\eqref{e27} for the scale factor, the effective and physical
energy densities, and the pressure after specifying an equation of
state for the effective and physical ideal fluids. Finally, we
should emphasize one crucial fact here, namely, that the
functional form of the matrix $\mathcal{V}^{\prime}$ is different
than that in Ref. \cite{massgrav}. This is a consequence of the
different solution ansatz used in the present work. The
cosmological solutions (the background and physical metrics as
well as the St\"{u}ckelberg scalars) stemming from this ansatz are
new, both in their mathematical notion and their physical content
which we comment on in the Conclusion.
\section{Conclusion}
We have adopted the solution method developed in Refs.
\cite{cosmommgr,massgrav} to derive new cosmological solutions of
the ghost-free massive gravity theory. For an unspecified
background metric, and a FLRW physical metric scenario, we first
decoupled the St\"{u}ckelberg scalars from the metric sector of
the theory via a specially chosen ansatz which introduces an
effective cosmological constant and a fluid contribution of the
mass sector to the metric equation. Then, we found solutions to
the scalars and derived the background metric in terms of these
scalar solutions, the physical FLRW metric, and the properties of
the effective fluid so that they algebraically satisfy the
solution ansatz. Later, we constructed the cosmological equations
where the collective effect of the St\"{u}ckelberg scalars and the
background metric enter into the equations as an effective
cosmological constant and a source in the form of the energy
density and the pressure of a hypothetical effective ideal fluid.
Therefore, after specifying a freely chosen equation of state for
the effective fluid one may first solve the scale factor of the
FLRW metric, the energy density, and the pressure of the effective
fluid and then use these to construct the necessary background
metric for which these solutions exist. Consequently, we have
obtained a large class of cosmological solutions of the theory for
diagonal-background metrics, where the diagonality has been the
key ingredient to decouple the components of the ansatz equation
(which is a matrix equation). The derived solutions compose a wide
class because one has solution-parametrizing functional degrees of
freedom in assigning a generic equation of state to the effective
fluid. In addition, the solutions are also parametrized by two
free coefficients. We should declare a fundamental difference
between the outcome of the present work and that in Ref.
\cite{massgrav}, both of which use the same solution systematics.
First of all, the solution ansatz and the resulting solutions are
quite different for these two works. The basic reason for this
lies in the nature of the ansatz-generating effective fluid. As we
have mentioned before, the solutions obtained in Ref.
\cite{massgrav} were parametrized by an effective ideal fluid
which has a modified energy-momentum conservation law. This fact
has structural influences not only on the solution ansatz, but on
the cosmological dynamics as well. All the cosmological equations
are modified due to the nonphysical nature of the effective fluid
in Ref. \cite{massgrav}. On the other hand, in the present work
where we found a new class of solutions of the theory the
effective fluid is assumed to obey a physical energy-momentum
conservation equation. This is reflected in two places: first, in
the simplicity of the solution ansatz, and second, in the
unmodified canonical form of the Friedmann equation and the
continuity equations of the cosmological dynamics. We believe that
the second of these is quite important. Fundamentally, reviving
the canonical form of the cosmological dynamics as solutions
within the context of massive gravity with just the addition of
some hypothetical matter is the same as restoring Einsteinian
cosmology in massive gravity. Hence, the formalism and the results
of the present work enable us to construct an inclusion map
between the homogeneous and isotropic solutions of Einstein
gravity and those in ghost-free massive gravity. One lacks such a
connection for the other branch of the solution moduli constructed
in Ref. \cite{massgrav}. This is not only an essential
mathematical achievement but it also promises physically eligible
solutions of massive gravity that may hold a cure for the dark
energy problem of standard cosmology in the effective-matter
domain. More specifically, the ordinary Einstein form of the
cosmological dynamics enables one to combine the physical and
effective-matter terms. This may help one to construct solutions
in which the geometry of the solutions does not match with the
required physical matter content (from an Einstein point of view),
where the effective matter takes the role of compensation. We
believe that such a significant result is the most outstanding
contribution of the present work and deserves to be reported in
its own right.

We should furthermore note that unlike the general achievements in
the literature the cosmological solutions derived here cover a
generic effective energy-momentum contribution in the cosmological
dynamics rather than just a cosmological constant one. Moreover,
the St\"{u}ckelberg field solutions are not restricted to a
particular functional set of solutions, as was the case in other
similar works. Rather they are completely arbitrary in the
solution framework we have addressed. As an example, the reader
may compare the method applied here with the one used in Ref.
\cite{ih1}, where it was proven that self-accelerating solutions
exist for an isotropic background metric whose functional
dependence on coordinates differs from the single-coordinate
dependence assumed in this work. One realizes that therein, for a
special choice of the background metric ansatz in terms of the
St\"{u}ckelberg fields, it has been shown that for a particular
configuration of solutions that fixes the scalar fields (which
depend on the scale factor unlike our case, also without exactly
being solved) thus the background metric the effective
contribution in the metric sector becomes a cosmological constant
hence giving rise to self-accelerating cosmological solutions on
the physical metric side. Such a comparison may reveal the
extensive generality of the solution domain obtained here not only
for the scalars and background metrics but also for the
cosmological scenarios, due to the richness of their effective
source feed originating from the large solution moduli of mass
degrees of freedom. The reader should also appreciate that even
trivial choices of St\"{u}ckelberg field solutions are possible in
this solution-generating method. The fundamental reason for this
extension of freedom is that we are able to design the background
metric for a desired set of scalar and physical metric solutions.
On the other hand, the price we pay for this is the nonstandard
and complicated form of the background metric.

 Similar to the structure of the solutions
in Ref. \cite{massgrav} the freedom of choice of the equation of
state of the effective matter enables us to construct a rich class
of new cosmological-solution scenarios of the ghost-free massive
gravity, which are certainly distinct from the ones obtained in
Ref. \cite{massgrav} both in their mathematical structures and
their physical nature. In particular, this broad class of
solutions contains the self-accelerating ones. For example, as a
special case all the solutions with $C_2=0$ are self-accelerating
as the overall contribution to the metric sector in this case
becomes an effective cosmological constant. We leave a detailed
study and classification of these solutions to a later work. We
should also note that the general application domain of the
solution method of this work is not restricted to the cosmological
cases only and it can be used to derive other solutions of the
theory. Besides even though we have assumed diagonality and a
special form for the background metric, one may work out the
general solutions of the ansatz equation without assuming a
particular form. This would result in a system of coupled
first-order partial differential equations but in this case the
important achievement would be to obtain the most general set of
FLRW solutions of massive gravity. Finally, we point out that a
future direction could be to inspect how to accommodate the class
of solutions presented here in the context of bigravity.
\section*{Acknowledgements}
We thank Merete Lillemark for useful communications.

\end{document}